\documentclass[aps,twocolumn]{revtex4-1}
\usepackage{amsmath,amssymb}
\usepackage{slashed}
\usepackage{graphicx}
\usepackage{bm}
\usepackage[T1]{fontenc}
\usepackage[utf8]{inputenc}
\usepackage{gauss} 
\usepackage[top=1.0in,bottom=1.0in,left=1.0in,right=1.0in]{geometry}
\usepackage[colorlinks,linkcolor=blue,citecolor=blue]{hyperref}
\newcommand{\be}{\begin{equation}}
\newcommand{\ee}{\end{equation}}
\newcommand{\bea}{\begin{eqnarray}}
\newcommand{\eea}{\end{eqnarray}}
\newcommand{\ba}{\begin{eqnarray}}
\newcommand{\ea}{\end{eqnarray}}

\begin{document}

\title{A cosmological ``Big Storm  Scenario" \\
following the  QCD phase transition
}

\author{Edward Shuryak}
\email{edward.shuryak@stonybrook.edu}
\affiliation{Center for Nuclear Theory, Department of Physics and Astronomy, Stony Brook University, Stony Brook, New York 11794--3800, USA}

\author{Gregory Falkovich}
\email{gregory.falkovich@weizmann.ac.il}
\affiliation{Weizmann Institute of Science, Rehovot 7610001 Israel }

\begin{abstract} 
It was proposed that acoustic perturbations generated by the QCD phase transition could create an inverse turbulent cascade \cite{Kalaydzhyan:2014wca}. An assumption was that propagation toward smaller momenta could reach 
the wavelength of a few km, the Universe's size at the time.
Such acoustic waves were proposed to be the source of gravity waves. The kilometer wavelength corresponds to year-long gravity wave today, which were recently discovered using pulsar correlations. 

This paper argues further that an acoustic turbulence must be an ensemble of shocks. This brings two consequences: First, shocks generate gravity waves much more efficiently than sound waves due to the intermittency of matter distribution. We reconsider gravity wave radiation, using universal emission theory for soft radiation, and argue that soft-momenta plateau should reach wavelengths of the order of shock mean free paths. Second,  collisions of shock waves create local
density excess, which may create primordial black holes. A good tool to settle this issue can be an evaluation of the {\em trapped surfaces} like it was done
in studies of heavy ion collisions using AdS/CFT correspondence.
 \end{abstract}

\maketitle
\section{Introduction}
The time of the cosmological phase transition follows from the standard Friedmann  solution for radiation-dominated expansion 
 \be \label{eqn_t_QCD_2}
 t_{QCD} =\big({90 \over 32 \pi^3 N_{DOF}}\big)^{1/2}{M_P \over T_{QCD}^2} \ee
 where $M_P=(\hbar c/G_N)^{1/2}$ is the Plank mass, corresponding temperature is $T_{QCD}\approx 155 \, MeV$ and $N_{DOF}$ is the effective number of bosonic degrees of freedom. 
 The corresponding length scale is  \be h_{QCD}\sim c\cdot t_{QCD}\sim 10\, km \label{eqn_t_QCD}\ee 

The scenario to be discussed originates from paper \cite{Kalaydzhyan:2014wca} which  can be briefly summarized as follows. \\
(i) Disturbances created at the cosmological QCD phase transitions 
produce  sound waves originating
at the (UV) scale $k\sim T_c$  ;\\
(ii) Subsequent acoustic turbulence enters an {\em inverse cascade regime} and as a result amplitude of sounds $n_k$ increases
as they propagate toward small momenta, $k\rightarrow 1/h_{QCD}$. The amplitudes of the waves increase, thus 
 the storm. The 
wavelength $h_{QCD}$ is multiplied by corresponding Universe expansion $z$ factor and constitute about one light
year today.
\\
(iii) The production of gravitational waves was described by the elementary process
{\em two sound waves -> gravity wave (GW)}. 

%

In the intervening years, several experimental discoveries and theoretical developments provided new reasons to focus on the Universe at the scale $h_{QCD}$.
As usual, new developments lead to many interesting questions and problems.
We will briefly review some of them below, which could be useful, as they touch on quite different fields of physics.

In the intervening years, the most important for the subject in question is that GW with period $\sim 1 \, year$ were in fact observed.   Several Pulsar Timing Array groups have reported the discovery of a stochastic gravitational wave background (SGWB). In particular, the North American Nanohertz Observatory for Gravitational Waves (NANOGrav) \cite{2306.16213}, the European PTA \cite{2306.16214}, the Parkes PTA \cite{2306.16215} and the Chinese CPTA \cite{2306.16216} have all released results which seem to be consistent with expected pattern in the angular correlations characteristic of the SGWB.
If these GW are of cosmological origin, one now gets the absolute scale of 
their amplitude at $t_{QCD}$, which can be compared to proposed mechanisms
of their radiation.
 
In this paper, we reconsider and significantly change the scenario, taking into account the fact that acoustic waves of finite amplitude in a medium with a low viscosity inevitably turn into shocks.

 We also propose that density perturbations induced by storms in cosmological settings
may create, under certain conditions, ``overdense" regions to undergo gravitational collapse,  generating ``primordial black holes", PBHs. Well-known studies of the phenomenon were all done in spherically symmetric density fluctuations in a standard
expanding Universe.  Studies of colliding shocks were done in
a completely different field of physics, describing ultrarelativistic heavy ion
collisions via AdS/CFT correspondence. 
BH formation has been studied by calculating trapped surfaces. There, general relativity was used
in higher dimensions, as dual to QCD, but methodically it is very close to what we need. This theory can easily be modified to address the collapse in the standard General Relativity. An important tool used in those papers is
evaluating the {\em trapped surfaces}, providing lower bounds on BH masses.

\section{Soft graviton emission without a  storm}

Let us start with a discussion of why there can be sounds after
the phase transitions, and whether they can go into a turbulent regime.

QCD phase transition 
 has been studied experimentally in heavy ion collisions, in the last decades, mostly at two colliders, RHIC at Brookhaven National Laboratory, USA and LHC at CERN. They study the transition of matter from the Quark-Gluon Plasma (QGP) phase to Hadron Resonance Gas (HRG).
 Relevant for our discussion are
studies of sound excitations. This is done via the so called azimuthal harmonics of the flow $v_m$ with the range $m=2...10$. They are sounds with the wavelength $\lambda_m=2\pi R_{fireball} /m$, ranging from $20$ to $4 \, fm$. Their amplitudes depend on sound velocity $c_s =dp/d\rho\approx 0.16$. Their
damping reveals surprisingly low viscosity of
the plasma.

 Naively, one may expect that the Universe expansion is rather slow, with the system having more than enough time to adjust to changing temperature $T(t)$ and be near thermal equilibrium. 
The phase transitions are still expected to produce perhaps certain out-of-equilibrium excitations:
the first-order transitions with its bubbles is the best-known example. ``Critical opalescence" while passing the second-order transitions is another well-known example. (A search for such a hypothetical QCD critical point is an experimental program at RHIC, so far without a widely accepted conclusion.) 
Small but finite quark masses smoothed the (chiral symmetry breaking) QCD transition at $T_c$,  believed to be just a crossover. 
Yet, even for such a regime, one may expect an inhomogeneous distribution of both phases.  
One particular model has been discussed in \cite{Shuryak:2013uaa}: instead of slowly evaporating, the QGP clusters should undergo collapse, transferring (part of) their energy/entropy into the outgoing sounds.  Instead of
bubble collisions, they individually participate in spherical collapse, as has
been demonstrated for bubbles in water by Rayleigh.  If so, most sounds
is emitted at wavelength corresponding to the smallest size of the collapsing QGP
cluster, perhaps again at the micro or UV scale $1/T_c$. 

In \cite{Kalaydzhyan:2014wca}, it was proposed that the collision of sound waves
can be the origin of GWs. Considering
 sound momenta $\vec p_1,\vec p_2$  and $\vec q$ of the GW to
be comparable, it was argued for a peak in the emission spectrum \cite{Kalaydzhyan:2014wca}. 
In this work we will focus instead on soft gravitons, with $q \ll p_1,p_2$.

The emission of soft gravitons, like that of soft photons,
follow particle collisions in universal factorized form pointed out by Weinberg
\cite{Weinberg:1965nx}, with extra factors in each line of the Feynman diagram  being
\ba f_{photons} &=& \sum_n {e_n \eta_n p_n^\mu \over (p q) -i \epsilon \eta_n} \\
f_{gravitons} &=& (8\pi G_N)^{1/2}\sum_n {\eta_n p_n^\mu p_n^\mu \over (p q) -i \epsilon \eta_n}
\nonumber
\ea 
where $\eta_n=1$ for outgoing lines and $-1$ for in-going ones,  $p^\mu, q^\mu$ are momenta of the parent particle and GW, respectively. The emission cross-sections are
singular at $q\rightarrow 0$, but only because they count a number of photons/gravitons. To give the power spectrum, the cross sections need to be multiplied by the energy $q_0=\hbar \omega$, then $\hbar$ cancels out in the dimensionless factors such as 
\be \label{eqn_GN} {G_N M_N^2 \over \hbar c}\approx 5.8\cdot 10^{-39} \ee
(with $M_N$ here being the nucleon's mass).  The resulting expressions for power production
has no $\hbar$, so they are valid in both quantum and classical domains.  

 Note that Weinberg's calculation contains the {\it transport} cross-section $\sigma_t$, namely the one weighted with the momentum transfer.
Let us explain why this is the case. The emitter of soft gravity waves summing over
incoming and outgoing lines for $2\rightarrow 2$ collisions  reads
\ba &-&p_{1,in}^\mu p_{1,in}^\nu -p_{2,in}^\mu p_{2,in}^\nu
+p_{1,out}^\mu p_{1,out}^\nu +p_{1,out}^\mu p_{1,out}^\nu \nonumber \\
&=&
 p_{1,out}^\mu p_{2,out}^\nu - p_{1,in}^\mu p_{2,in}^\nu +(\mu\leftrightarrow \nu)
\ea
where the second line is obtained using
momentum conservation $p_{1,in}^\mu+p_{2,in}^\mu=p_{1,out}^\mu+p_{2,out}^\mu$. It vanishes for forward scattering, $p_{1,in}=p_{1,out},p_{2,in}=p_{2,out}$.
So, there is no soft GW radiation without a change in the stress tensor $T^{\mu\nu}$.

For nonrelativistic collisions producing gravitons
with energy below $\Lambda$, Weinberg calculated the
total power emitted via soft gravitons to be
\be  P(q_0<\Lambda)= {8 G_N \Lambda \over 5 \pi} m^2 v^5 n_a n_b \sigma_{transport} V 
\ee
where $n_a,n_b$ are densities of colliding particles, $V$ is interacting volume, $v$ velocity and $\sigma_{transport}$ 
is the transport cross section (weighted with $sin(\theta)^2$). As an exercise, for electron-proton collisions in the Sun, he estimated this power. Substituting regulated cross section one gets the power of the Sun about $P_{Sun\, gravitons}\approx 7*10^7 \,W$ emitted in sub-$keV$
gravitons. For comparison, total Sun power in thermal photons is $P_{Sun}=3.8\cdot 10^{26}\, W$,
and comparable in solar neutrinos. 

(A reader may wonder why GW radiation is  only 19 orders of magnitude smaller
than electromagnetic, while its coupling is smaller than $e^2/\hbar c=1/137$ by about 38 orders of magnitude? 
Without going into complicated nuclear physics of the Sun energy production, 
let us just note one effect:  soft photons are emitted into
a heat bath in thermal equilibrium, with emission 
 balanced by absorption. GW radiation, on the other hand, just leave unobstructed.)

Let us now evaluate  production of soft gravitons accompanying hadronic collisions in hadron resonance gas (HRG) after the QCD phase transition. In the plasma phase, effective masses of quark and
gluon quasiparticles are $O(1\, GeV)$. Most hadronic species in HRG have the same order of magnitude. Comparing to typical thermal momenta $3T_{QCD}\sim 0.5\, GeV$ one may take the nonrelativistic estimate as 
qualitatively valid. With $M=1\, GeV, n_a=n_b=1/fm^3, V=1\cdot m^3$ we estimate the soft graviton
production to have power per $dV dt$ 
$$P_{HRG}\sim 6\cdot 10^{18} kg/m/s^3$$
The energy of GW produced during time $\sim 10t_{QCD}\sim 10^{-4} s$
is accumulated gives the achieved energy density (per $m^3$)
$$\rho_{gravitons}\sim 6\cdot  10^{14} {kg \over m s^2}$$
For comparison, order of magnitude of the energy density of hadronic matter itself is
of the order of 
$$\rho_{hadrons}\sim 1\, GeV/fm^3\sim 1.6\cdot 10^{35} {kg \over m s^2}$$
is 21 orders of magnitude larger. 

(Note that small radiation rate, containing a very small dimensionless ratio (\ref{eqn_GN}), is somewhat compensated
by the relatively long duration of the emission time. Gravitons, unlike photons and strongly interacting species, are accumulated during emission time, with negligible absorption. This type of enhancement of any ``penetrating probe" production is well known in the heavy-ion collision community, among others.
A more formal answer is to note that the duration
$t_{QCD}$ (\ref{eqn_t_QCD_2}) is proportional to the Plank mass
$M_P=\sqrt{\hbar \cdot c /G_N}$, so the total energy density of the soft GW produced
is in fact proportional to a square root $\sqrt{G_N}$, not the coupling $G_N$ itself.) 

In recent studies of soft GW production from
cosmological perturbations \cite{Sharma:2023mao}
 a universal shoulder in spectrum left of  the acoustic peak was observed, called ``shallow relic" by the authors.   Its energy spectrum is not constant but linear in momentum
$\Omega_{GW} \sim k $: the difference with our shoulder  is because their model is quite different, so
 there is no contradiction between both results.

After our simiplest estimates, let us jump
to pulsar data and see how much GW power is needed to reproduce them. The observed radiation in
energy density units now can be converted to the phase transition time $t_{QCD}$ by energy density scaling $z^4$.
The corresponding factor $z_{QCD}=T_{QCD}/T_{now}\approx 6\cdot 10^{11}$.
Using observed 
\be  \rho_{GW\,now}\approx 7.1\cdot 10^{-17} {kg \over m s^2}
\ee
we get the energy density at the time of the QCD transition 
\be \rho_{GW\,then}\approx  9 \cdot 10^{30}  {kg \over m s^2} \ee
This is only $10^{-5}$ of matter density, but many orders of magnitude larger than due to
soft graviton emission from hadron collisions evaluated above. Also of course
the frequency do not match. So, the simplest source of GW production -- hadron collisions --
fails badly. 

A ``storm" idea substitute microscopic colliding objects (hadrons) by macroscopic ones, the ``pancakes" of shocks, so masses and cross sections are greatly increased. The shoulder may reach correct wavelengths 
of $km$. At this time we cannot provide quantitative estimates of those, so we do not yet know if this mechanism 
can or cannot explain the observations. 

\section{Acoustic turbulence as an ensemble of  (relativistic) shocks}

\begin{figure}
    \centering
\includegraphics[width=8cm]{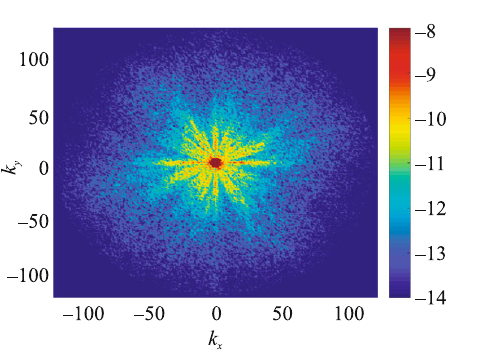}
    \caption{
    The energy spectrum $E_{\vec k}$ in momentum space. The peaks correspond to ``pancake-shaped" shock waves. 
    Their height is $\sim 1/d$, and the width $\sim 1/R$, where $d,R$ are the width and radius of these pancakes.
    }
    \label{fig_Kuz}
\end{figure}

The turbulence of acoustic waves in a medium with sufficiently low viscosity leads to a cascade of sound waves, eventually evolving into a set of shock waves, see e.g.  Fig.\ref{fig_Kuz} taken from recent numerical simulation \cite{KochurinKuznetsov}. 
In our case of  QGP/hadronic gas, the velocity of even weak shocks coincides with the speed of sound $c_s=c/\sqrt{3} \approx 0.55\cdot c$, and is comparable to the speed of light. Stronger shocks move even with a larger rapidities. Therefore, one must use relativistic notations we are going to remind now. 
Hydrodynamics is about motion of matter defined via energy density $\rho$, pressure $p$, and collective flow velocity vector $v_i,i=1,2,3$. Two former
quantities are Lorentz scalar, and  velocity  is promoted by 4-vector $u^\mu,\mu=0,1,2,3$.  The stress tensor in local approximation is
\be T_{\mu\nu}=(\rho+p)u^\mu u^\nu +p g^{\mu\nu } \ee
Consider motion  in one dimension, so there are two components of 4-velocity $u^0,u^1$, 
 constrained by $(u^0)^2-(u^1)^2=1$. Using
$u^0=cosh(y),u^1=sinh(y)$ one introduced the parameter $y$
called $rapidity$. Unlike velocity, its Lorentz transformation is done by a simple shift.

\begin{figure}
\includegraphics[width=6cm]{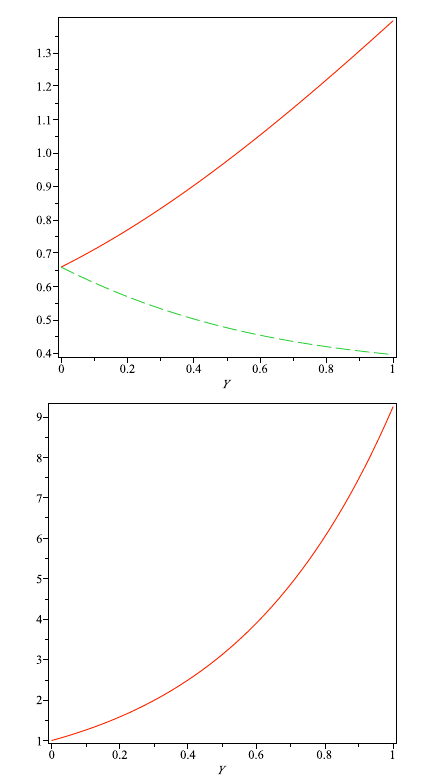}
\caption{(Upper plot) The rapidity of the shock $y$ versus the matter rapidity $Y$ is shown by upper (red) solid line, while the lower (green) dashed line shows rapidity of another side $y -Y$. (Lower plot) The compression ratio of matter versus $Y$ .}
\label{fig_rel_shock}
\end{figure}

The shock relations
are given by continuity of energy and momentum flow.
Omitting details (see \cite{Shuryak:2012sf} ) we explain the setting and results. Consider relativistic matter moving with
rapidity $y$ and undergoing energy density jump at a shock moving with rapidity
$Y$. After becoming compressed, the matter has rapidity $y-Y$.
The solution of the continuity equations is as follows:
\be y(Y)=Y-{1\over 2}log\big( e^{2Y}+1-\sqrt{e^{4Y}+e^{2Y}+1}\big)\,.\ee
Fig.\ref{fig_rel_shock} shows the relation 
 for a relativistic fluid ($p=\rho/3$).
As you can see when the shock is weak ($Y\approx 0$), 
matter compression is small, and its speed is that of the sound  $y=tanh(c_s)=tanh(\sqrt{1/3})$. 

With increasing $Y$ the compression contrast between upstream and downstream matter grows, reaching about an order of magnitude for $Y\sim 1$.
As a numerical example to be mentioned below, for $Y\approx 1/2$ the shock rapidity $y\approx 1$ and velocity $v_{shock}=tanh(y)\approx 0.76$, and compression $\sim 2.5$.

\section{Gravity wave emission from shock ensemble}

Standard approaches to GW emission are based on the correlator of two
 stress tensors. For an emission by sound waves of small amplitude, the terms $(\rho+p) u^\mu u^\nu  $ are expanded in the wave amplitudes. Since the velocity is of the first order, $(\rho+p)$ is taken in the zeroth order.  The correlators of velocities are usually
calculated assuming an independent Gaussian ensemble of waves. The mean power of gravity waves emitted by sound is proportional to the mean squared power of acoustic waves, which was estimated as the squared mean power.   

Such estimates are off by a large factor for a highly intermittent set of shock waves. Indeed, for a set of locally plane shocks, one can use the Burgers scaling for the moments of the velocity difference $\delta v$ taken at the distance $l$: $\langle  (\delta \vec v)^n \rangle\simeq v_{rms}^nl/L$, where $L$ is the mean distance between shocks. That gives   the fourth moment of the velocity difference strongly enhanced in comparison with the squared second moment
\be 
\langle  (\delta \vec v)^4 \rangle\simeq   \langle (\delta \vec v)^2\rangle  ^2{L\over l}\,.
\ee
The largest enhancement is for $l\simeq d$, where $d$ is the width of the shock front.
Furthermore, in strong turbulence with shocks, there is no small-amplitude expansion since $\delta (\rho+p)$ and 4-velocities $u^\mu$  are 
not small but rather $O(1)$.

Both nonrelativistic and relativistic shocks have the same 
 characteristic feature: stronger
ones move quicker, catching and absorbing weaker ones so shock turbulence gets stronger with time. So, suppose the main events happening are $2\rightarrow 1$ processes. Let momentum conservation be expressed as  $p_1^\mu+p_2^\mu=P^\mu  $. The soft gravitons are then emitted by the stress tensor change
\ba \delta T^{\mu\nu}&=&P^\mu P^\nu-p_1^\mu p_1^\nu-p_2^\mu p_2^\nu \nonumber \\
&=& p_1^\mu p_2^\nu+p_1^\nu p_2^\mu
\ea
which in this case is always nonzero. Therefore emission rate
should include the ordinary cross-section (rather than the transport one, as was the case for 
the $2\rightarrow 2$ scattering).

Let us now clarify the following question: How can soft gravitons be emitted at large times/distances between shock collisions when they do not yet overlap? Let us start answering it with soft photons. While a charged particle can be well localized on its trajectory, its e/m field cannot. In particular, at rest, it is a Coulomb field $A_0\sim e/r$.

\begin{figure}
\includegraphics[width=8cm]{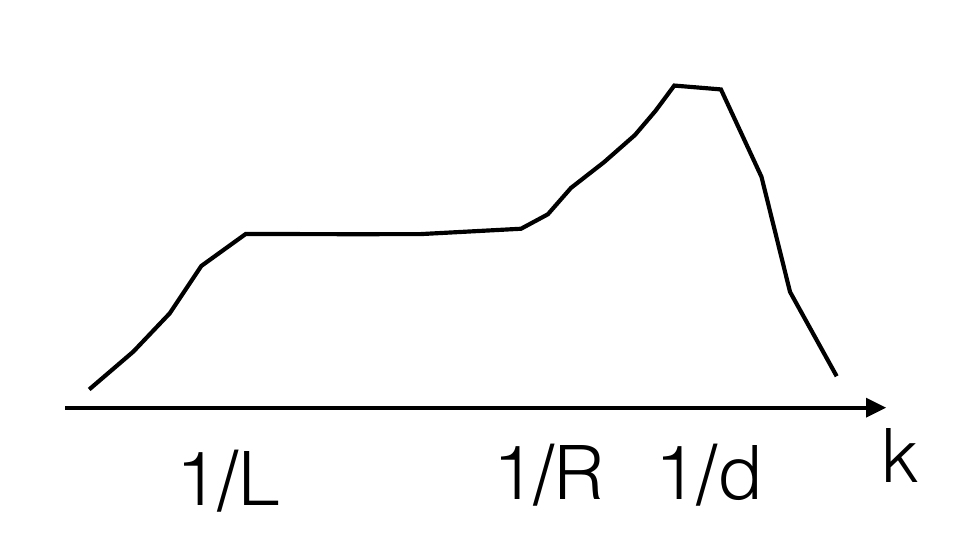}
\caption{Schematic shape of the GW energy spectrum emitted by shock ensemble. For the meaning of the indicated parameters, see text.}
\label{fig_sketch}
\end{figure}

If objects are locally flat ``pancakes", the transverse size $R$ is much larger than the width $d$.
Face-to-face collisions will have the shortest
duration $t_{face-to-face}\sim d/cosh(Y)$ ($Y$ is shock rapidity, cosh gives compression by the Lorentz factor). Collisions of ``pancakes" with nonzero
angles last longer, all the way to
orthogonal collisions with  $t_{angle}\propto R$.  Radiation emitted would have an asymmetric peak at the corresponding frequencies, ranging from  $1/t_{face-to-face}$ to $1/R$.
A schematic picture of the gravity wave spectrum is shown in Fig.\ref{fig_sketch}. Explicit calculations for 
various models of turbulence after the phase transitions
(such as \cite{Sharma:2023mao})  lead to similar shapes.

We claim that a universal shoulder in the spectrum toward the IR is due to the emission of soft gravitons. As we discussed above, it is
due to general expressions for soft emissions from collision processes in which the stress tensor suffers rapid changes. 

An important question is how far the soft GW shoulder reaches toward small $k$. Note 
that their emission happens during {\em the time between collisions}.
As pointed out by Feinberg and Pomeranchuk in 1950's, after each scattering, the electron appears ``semi-naked", with a reduced electric field, which gradually regenerates later. Obviously, the same should be true for soft GW.

If shocks move relativistically,  their gravity field also gets Lorentz contracted into a  pancake. When they change velocity or direction of motion, part of their Newtonian field continues into the old
direction, constituting the soft radiation we discuss.
We, therefore, state that the universal shoulder should reach the mean free path
of the radiated objects. For the ensemble of shocks 
 this longest time scale is  
\be L\sim {1 \over n \sigma v}\sim  {1 \over n_{shocks}(4\pi R^2)}\ee

Furthermore, since we are aiming to explain the observed GW background with periods $\sim 1$ year today, or  $\lambda \sim c t_{QCD}\sim km $, then the left end of this shoulder should reach this scale, which is many orders larger than the microscopic scales defined by the ambient temperature $\lambda \sim 1/T_{QCD}$, with quasiparticle/hadrons mean free paths perhaps one or two orders of magnitude longer.  Our proposal is that the strong turbulence leading to
a picture of a rather dilute ensemble of shocks may eventually reach this scale.
This implies that the shock density must be smaller than
\be n_{shocks} < {1 \over c \cdot t_{QCD} \cdot (4\pi R^2)}  
\ee
where, we remind, $R$ is the transverse size of shock ``pancakes". 

\section{Primordial black holes from shock collisions?} \label{sec_PBH}
Relativity and astronomy textbooks explain how BH may be produced in conventional stellar collapse. For decades, astronomers observed Supernovae without apparently nothing visible left at the center.
Nowadays, after the LIGO/VERGO collaboration \cite{Ricci:2022kut} 
have observed BH mergers, they start getting some clues about BH mass spectrum.
A significant  surprise from the first event is that it is not easy to explain how  BH with masses $M_{BH}\sim 20-30\, M_{Sun} $ were produced.

 People wondered if those could be primordial ones (PBHs).  
Unfortunately,  very little is known about the mass/spin distribution of BHs.
The first observation of isolated BH nearby was recently detected by microlensing \cite{OGLE:2022gdj}. Its mass is about $7\, M_{Sun}$, so it
can well be of stellar origin. Interesting that distance is close enough to use the classical parallax method: so perhaps their density is rather high. 

In principle, one may be able to tell primordial black holes (PBHs) from those created in stellar collapse by different momentum and spin distributions. Stellar BHs are expected to have large momentum due to
``natal kicks" from supernovae, producing  distinct distribution 
in Galaxy gravity field, e.g. be somewhat away from the
Galaxy plane in which stars are.  
 Perhaps with time, more will be found, and some ideas about their masses and space distribution of BHs will be clarified.   

The idea that energy density fluctuations in the early Universe can lead to PBH originated by Zeldovich and Novikov \cite{Novikov:1965sik} and then Hawking and especially Carr, see e.g. \cite{Carr:1975qj}.
While involved in Universe expansion, such matter as QGP or hadron resonance gas has pressure large enough to withstand 
gravitational instabilities. Yet, if some
subvolume happens to have sufficient excess density, it may collapse. 
Early estimates \cite{Carr:1975qj} of the critical values
\be {\delta \rho\over \rho} > \delta_c \sim c_s^2={dp \over d\rho} \ee
needed for that relate it to the speed of sound. Near the QCD
transition temperature, $c_s$ has a minimum known in heavy ion physics as ``the softest point" of the equation of state.
 It was studied hydrodynamically \cite{Hung:1994eq} and then experimentally observed
 via a peak in the fireball lifetime as a function of the collision energy of heavy ions. Therefore, a minimum in the speed of sound, located in the vicinity of the QCD phase transition,  should be suspected location for 
 gravitational collapse.
  It is straightforward
 to estimate BHs masses produced, as they should contain all matter
 in a cell of size (\ref{eqn_t_QCD}). This leads to  $$M_{PBH}\sim 20\, M_{Sun}$$ 
which is exactly in the ballpark of puzzling BHs seen by LIGO in mergers. Whether PBHs can or cannot be a significant fraction of the {em dark matter} is a hotly debated subject. Let us just comment that even if it is, than PBHs matter
fraction at the QCD transition time should be of the order of only $\sim 10^{-9}$.

Multiple numerical solutions of Einstein's equations has been made over the years to detail the process. As an example, consider \cite{Musco_2005} and references therein. As in other works, spherical symmetry was assumed, with bulk matter flowing according to the usual (radiation-dominated) Friedmann solution. 
If density is supercritical, with time, one finds matter cells with positive and negative (directed inside) radial velocities, separated by a ``void".
(An interesting ``second falling shell of matter" was also observed in these calculations.) 
The critical density contrast was found to be $\delta_c=\delta \rho/\rho\sim 0.43-0.47$. If
 density just slightly exceeds it, the mass of the produced PBS
gets to be much smaller 
\be {M_{PBH} \over M_H}\sim (\delta-\delta_c)^\gamma \ee
traced over several orders of magnitude toward $10^{-4}$  or so, with  the 
 index  $\gamma\approx 0.36$. 
An interesting option, then, is to
 search for BHs with masses $smaller$ than star collapse can generate. So far, none of such have been seen in mergers. In fact, there seems to be an empty gap in the LIGO data, between 2 and 5 $M_{Sun}$.

Returning to the scenario under consideration, one may ask if colliding shocks may
 provide strong enough
 density perturbation for a gravitational collapse.  While locally the shocks themselves include significant matter compression, their sizes still are expected to be much smaller than visible size of the Universe, $d,R\ll c \cdot t_{QCD}$.
 On the other hand, their collisions
 include  matter going toward the center with relativistic rapidities, which
 undoubtedly helps to initiate the gravitational collapse.
 Corresponding calculations can be done by simulating Einstein equation in such geometry, which is a task for the future. 

Ultrarelativistic collisions in the hydrodynamical framework were pioneered by 
Landau \cite{Landau:1953wku}, who applied hydrodynamics to ultrarelativistic collisions of hadrons and nuclei. Due to Lorentz contraction, those look like ``pancakes of matter". While considered crazy at the time of writing, it described well the first rapidity distributions from 
colliders  \cite{Shuryak:1972zq}.
Now, the relativistic hydrodynamics does provide a description of heavy ion collisions
in many details, see e.g. \cite{Shuryak:2014zxa}.

A revival of studies of ultrarelativistic collisions  happened in the framework of
AdS/CFT holographic correspondence, relating certain strongly coupled gauge theories to string theories in higher dimensions. In the so called Maldacena limit, the latter
  basically are Einstein's general relativity in 5 dimensions. 
Such correspondence reproduces many observed features of Quark-Gluon Plasma (QGP)  including the equation of state and viscosity. 
 Ultrarelativistic collisions of shock-like objects were studied numerically and analytically, see e.g. 
\cite{Chesler:2010bi} and subsequent literature.

\begin{figure}
\includegraphics[width=8.cm]{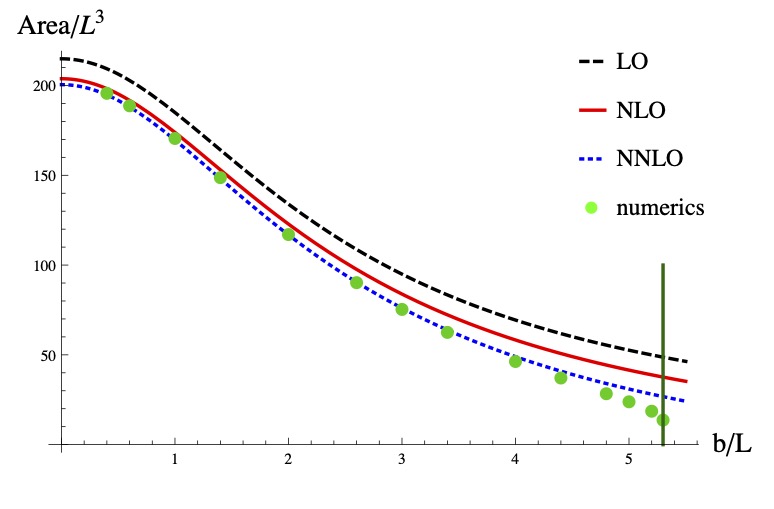}
\caption{Area of trapped surface -- the lower bound on the entropy of produced black hole -- as a function of impact parameter. Curves are for different approximations developed in  \cite{Gubser:2009sx}, and green points are from earlier numerical solutions in \cite{Lin:2009pn}. Vertical line at the right side
shows the impact parameter at which trapped surface disappears, indicating that BH production stops sharply at finite $b$.}
\label{fig_trapped}
\end{figure}

We want to emphasize one aspect of such calculations clarifying  
the production of the BHs in shock collisions.  When colliding objects
are about to overlap for the first time, so that the nontrivial solution 
of Einstein equations has not yet started, one can already study if
 the so-called {\em trapped surface} (for massless particles) may or may not be formed. 
 If it does, its area provides the lower bound for the BH entropy. 
 This tool was used in several publications a decade ago.
A very interesting feature of the trapped surface
(found numerically by one of us in \cite{Lin:2009pn}) is demonstrated in
Fig.\ref{fig_trapped} from \cite{Gubser:2009sx}. What is shown is the trapped surface area as a function of the impact parameter. At some critical value of the impact parameter, the trapped surface disappears completely while the area is still nonzero. That implies that mass distribution of produced BHs may have a sharp 
cutoff at some nonzero value.

 \section{Summary}
We have proposed a ``Big Storm Scenario" of what $may$ happen after the cosmological QCD
phase transition. The main new idea introduced here is that the transition produces acoustic turbulence, which naturally becomes an ensemble of colliding shocks absorbing each other, 
getting stronger and more dilute with time.

We discuss the mechanism of the soft graviton production from shock collisions. 
Specifically, we focus on  the smallest momenta, related to mean free paths of these shocks, which may reach the most interesting $km$ scale. If so, it may be the origin of the recently observed 
stochastic gravitational wave background of roughly a year period.
Another intriguing issue is whether these shock collisions can seed the Universe with primordial black holes. 

Many aspects of our scenario are hypothetical and suggested by and analogy to phenomena in related but different settings. 
Yet it is mostly based on ``known physics". Many assumptions  made can be 
clarified, by  well-defined 
numerical simulations. The properties of shock distributions can be found, and the details of black hole production in shock collisions can be studied.
We do understand EOS and many other properties of matter under consideration
from heavy ion experiments.
One cannot say that about alternative scenarios based on hypothetical fluctuations
of the inflation process.

{\bf Acknowledgements} We dedicate this paper to the memory of Vladimir E. Zakharov, a teacher, friend, and a giant of the theory of turbulence, who recently left us.\\
The work of ES  is supported by the Office of Science, U.S. Department of Energy under Contract No. DE-FG-88ER40388. The work of GF was supported by the Excellence Center at WIS, by the grants  662962 and 617006 of the Simons  Foundation,    and by the EU Horizon 2020 programme under the Marie Sklodowska-Curie grant agreements No 873028 and 823937. GF thanks  for hospitality the Simons Center for Geometry and Physics at Stony Brook.


\bibliography{BSS}

\end{document}